\documentclass[10pt,letterpaper]{article}
\usepackage{opex3}
\usepackage{cite}

\begin{document}

\title{Giant optical rotation\\
in a three-dimensional semiconductor\\
chiral photonic crystal}

\author{S. Takahashi,$^{1,\ast}$ A. Tandaechanurat,$^1$ R. Igusa,$^2$ Y. Ota,$^1$\\
J. Tatebayashi,$^1$ S. Iwamoto,$^{1,2}$ and Y. Arakawa$^{1,2}$}

\address{$^1$Institute of Nano Quantum Information Electronics,\\
University of Tokyo 4-6-1 Komaba, Meguro-ku, Tokyo 153-8505, Japan\\
$^2$Institute of Industrial Science,\\
University of Tokyo 4-6-1 Komaba, Meguro-ku, Tokyo 153-8505, Japan}

\email{$^{\ast}$shuntaka@iis.u-tokyo.ac.jp} 



\begin{abstract}
Optical rotation is experimentally demonstrated in a semiconductor-based three-dimensional chiral photonic crystal (PhC) at a telecommunication wavelength.
We design a rotationally-stacked woodpile PhC structure, where neighboring layers are rotated by $45^{\circ}$ and four layers construct a single helical unit.
The mirror-asymmetric PhC made from GaAs with sub-micron periodicity is fabricated by a micro-manipulation technique.
The linearly polarized light incident on the structure undergoes optical rotation during transmission.
The obtained results show good agreement with numerical simulations.
The measurement demonstrates the largest optical rotation angle as large as $\sim23^{\circ}$ at 1.3 $\mu$m wavelength for a single helical unit.
\end{abstract}



\section{Introduction}

Optical activity, comprised of optical rotation and circular dichroism on the basis of the circular polarization (CP), can be observed when light passes through a structure without mirror and spatial inversion symmetry, such as helices or chiral molecules.
Optical rotation is the rotation of the plane of linear polarization (LP) after transmission.
It is also known as circular birefringence because it is caused by the real part of the refractive index difference between the right-handed and the left-handed CPs.
On the other hand, circular dichroism originates from the difference in the imaginary part of the refractive index between the two orthogonal CPs, {\it i.e.} the difference of absorption between the two CPs.
Therefore, optical activity provides a way to control CP, which could be useful in potential applications including not only CP-light emitting devices\cite{Konishi}, but also quantum information processing in solid state systems.

Optical activity is small for the aforementioned applications in natural materials such as quartz, but can be enhanced in artificially-prepared chiral structures whose period is comparable to the wavelength of the incident light to be modulated.
A lot of works have recently been devoted to the fabrication of such chiral structures to realize large artificial optical activity and a high degree of polarization control.
Quasi-two-dimensional chiral systems made from metals, dielectrics, or semiconductors exhibit largely modified electric fields at the material interface, and show large degrees of optical rotation\cite{Gonokami,Konishi2,Konishi3}.
Three-dimensional (3D) chiral or helical structures are expected to further enhance the optical activity.
Several pioneering works using liquid crystals\cite{Coles}, polymers\cite{Thiel,Gu}, metals\cite{Lakhtakia,Plum,Zhao}, and dielectrics\cite{Lakhtakia} have been reported.
Semiconductor-based 3D chiral structures can be useful for applications such as highly efficient lasers emitting CP light and spin-photon interfaces\cite{Greve,Gao,Fujita,Kosaka}, since it is possible in semiconductor systems to easily introduce photon emitters and to manipulate electron/hole spins confined in nanostructures.
However, fabrication difficulty has largely limited the progress on semiconductor-based 3D chiral structures.
A Si-based 3D chiral structure has been fabricated and investigated for its 3D photonic bandgap, but not for optical activity\cite{John}.
Recent development of a micro-manipulation technique which enables the realization of high quality 3D photonic crystals (PhCs) based on GaAs\cite{Aoki,Aoki2,Aniwat} will overcome the difficulty.
Since this technique has large flexibility in various parameters, such as constituent materials and crystal structures, it is applicable to fabricate a 3D semiconductor chiral PhC with high quality and high flexibility.

In this study, we demonstrate for the first time a GaAs-based 3D chiral PhC.
We achieve the largest optical rotation angle of $\sim 23^{\circ}$ for a single helical unit at a telecommunication wavelength, which is, for example, 30 times larger than that in a twisted PhC fiber\cite{Russell}.
This will pave the way to various CP-based devices and applications.


\section{Design of a three-dimensional chiral photonic crystal}
A schematic of the studied 3D PhC is shown in Fig. 1(a).
In this article, we name our chiral PhC a rotationally-stacked woodpile structure.
The structure in air comprises a stack of dielectric thin layers with a thickness of 225 nm.
The first layer consists of periodic rods with 100 nm width separated by 500 nm.
The second layer placed on top of the first layer also has periodic rods with the same width, but the spacing is $500/{\sqrt2}$ nm and the pattern is rotated by $45^{\circ}$ with respect to the first layer.
The third (fourth) layer placed on top of the second (third) layer consists of the same periodic rods as those in the first (second) layer, but the rods are orthogonal to the first (second) layer.
A helical unit constructed by these four layers belongs to a spatial group with ${4_1}$ screw operation.
We use different spacing in the first (third) and second (fourth) layers to make crossing points aligned in the top view.
In this study, nine layers are stacked in total with a total thickness of $\sim$ 2 $\mu$m, and the pattern in the ninth layer is same as that in the first layer.
The $x$ ($y$) axis is defined to be orthogonal (parallel) to the rods in the first or ninth layer as shown in Fig. 1(a).

\begin{figure}[t]
\centering\includegraphics[width=9cm]{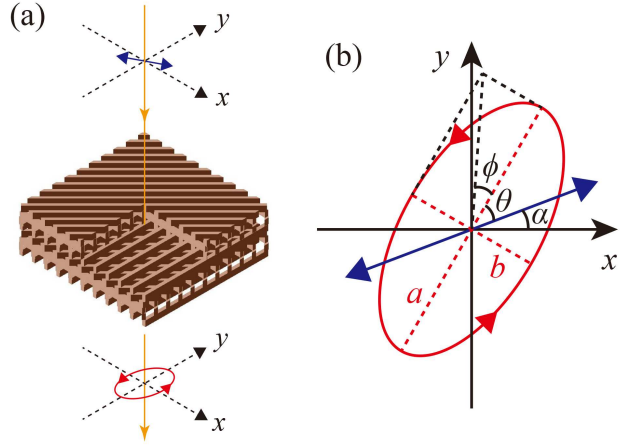}
\caption{
(a) A schematic illustration of the rotationally-stacked woodpile structure.
The top (ninth) layer is removed and a part of the structure is cut for clarity.
An impulse is irradiated normally to the structure.
Yellow arrows indicate the direction of the incident and transmitted light.
Blue linear and red circular arrows denote the polarization states of the incident and transmitted light, respectively.
(b) A schematic expression of polarization states (Blue: linearly polarized incident light, red: elliptically polarized transmitted light).
$\theta$ and $\phi$ are the polarization rotation angle and ellipticity.
}
\end{figure}

\begin{figure}[t]
\centering\includegraphics[width=12cm]{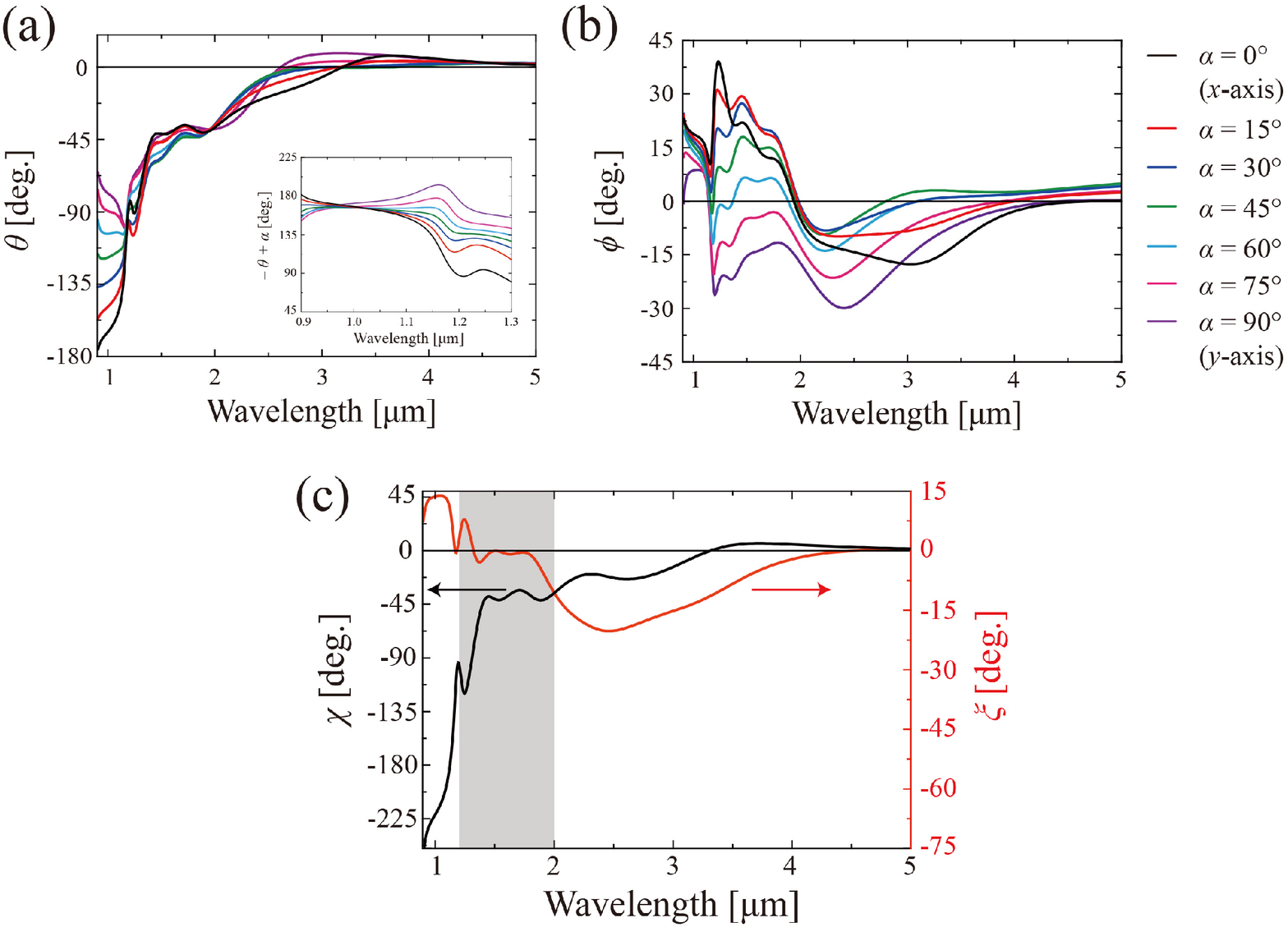}
\caption{
Calculated dispersion of the polarization rotation angle $\theta$ (a) and the ellipticity $\phi$ (b).
Inset in (a) shows a dispersion of $- \theta + \alpha$ in the region $\lambda < 1.2 \mu$m.
(c) The dispersion of the genuine optical rotation angle $\chi$ and the genuine ellipticity $\xi$ obtained by fitting the numerical data in (a) and (b) with the equation in the main text.
In the shaded wavelength region 1.2 $\mu$m $<$ $\lambda < 2 \mu$m, clear optical rotation is observed with finite $\chi$ and small $\xi$.
}
\end{figure}

To investigate optical activity in the chiral structure, we numerically simulate the polarization state of light transmitted through the structure using a finite-difference time-domain (FDTD) method.
The dielectric material is assumed to be GaAs with its refractive index $n$ = 3.4.
Periodic boundary conditions are imposed in both $x$ and $y$ directions.
The incident light is an impulse plane wave with LP, and enters through the top of the structure as shown in Fig. 1(a).
The time-dependent components of the transmitted electric field are recorded, and we analyze the spectral response by Fourier transforming them.

Suppose that the transmitted light has an elliptic polarization when light with a LP is irradiated as shown in Fig. 1(b).
Here, $\alpha$ is the azimuthal angle of LP in the incident light measured from the $x$ axis.
We define the polarization rotation angle $\theta$ as the angle between the LP direction of the incident light and the major axis of the ellipse in the elliptically polarized transmitted light.
Note that $\theta$ is not identical with an optical rotation angle, since the polarized plane of LP can be modified not only by optical rotation but also by linear birefringence and linear dichroism\cite{Konishi2}.
The genuine optical rotation angle $\chi$ can be derived by the $\alpha$ dependence of $\theta$ as discussed later.
We also define $\phi$ satisfying tan$\phi$ = $b/a$ as the ellipticity of the elliptically polarized transmitted light.
Here, $a$ and $b$ are the length of the major and minor axis of the ellipse, respectively.
The positive (negative) sign of $\phi$ indicates left-handed (right-handed) polarization.
$\phi$ also reflects not only circular dichroism but linear birefringence/dichroism, and is not identical with the genuine ellipticity $\xi$ which can also be derived by the $\alpha$ dependence of $\phi$.

Figures 2(a) and 2(b) show the numerically obtained dispersions of $\theta$ and $\phi$, respectively.
In the region of the incident wavelength $\lambda > 4 \mu$m, both $\theta$ and $\phi$ get close to zero with increasing wavelength, since light becomes insensitive to a structure much smaller than its wavelength.
At $\lambda < 1.2 \mu$m, $\theta$ largely depends on $\alpha$ and $\phi$ is almost independent on $\alpha$.
Further investigation shows that $-\theta+\alpha$, which is the angle between the major axis of the ellipse in the elliptically polarized transmitted light and the $x$ axis in the case of $\theta < 0$, is constant for various $\alpha$ as shown in the inset in Fig. 2(a).
This indicates that the structure acts as a polarizer which transmits only a particular polarization at $\lambda < 1.2 \mu$m.
In the wavelength region 1.2 $\mu$m $<$ $\lambda < 4 \mu$m, $|\theta|$ increases as the wavelength becomes shorter, with small dependence on $\alpha$, indicating optical rotation.
This slight $\alpha$ dependence of $\theta$ is due to linear birefringence/dichroism, which is also suggested by the $\alpha$ dependence of $\phi$ in Fig. 2(b) for 1.2 $\mu$m $<$ $\lambda < 2 \mu$m.
Different from $\theta$, the ellipticity crosses $\phi = 0^{\circ}$ and changes its sign for various $\alpha$ at 1.2 $\mu$m $<$ $\lambda < 2 \mu$m.
This indicates that the $\alpha$ dependence of $\phi$ is rarely caused by circular dichroism, but mainly caused by linear birefringence/dichroism.
The remaining linear birefringence/dichroism is mainly due to the pattern-induced in-plane index anisotropy.
Note that $\phi$ is negative for all $\alpha$ in most of the 2 $\mu$m $<$ $\lambda$ $<$ 4 $\mu$m range, indicating circular dichroism.

To distinguish the linear birefringence/dichroism effect from the optical rotation effect, we discuss $\theta$ as a function of $\alpha$.
While the optical rotation effect on $\theta$ is independent from $\alpha$, the linear birefringence/dichroism effect on $\theta$ depends on $\alpha$ with $\pi$ periodicity.
Hence, the linear birefringence/dichroism effect on $\theta$ can be expanded in a series of $\sin2\alpha$ as the following equation\cite{book,theory},
\begin{eqnarray}
\theta = \chi + \sum^{\infty}_{n = 1}A_n\sin(2n\alpha - \alpha_n).
\end{eqnarray}
Here, $\chi$ is the genuine optical rotation free from linear birefringence/dichroism, and $\alpha_n$ is phase offset for the $n$th order of $\sin2\alpha$.
Note that the top layer of the studied structure has in-plane anisotropy, leading to in-plane anisotropic reflection of the incident light.
This anisotropic reflection gives the same effect on $\theta$ as linear dichroism.
We found that taking the first two terms of the sum could adequately describe the simulated relation between $\theta$ and $\alpha$,
\begin{equation}
\theta = \chi + {\rm A_1}\sin(2\alpha - \alpha_1) + {\rm A_2}\sin(4\alpha - \alpha_2).
\end{equation}
Upon fitting the data for all wavelength, we could obtain the dispersion of the genuine optical rotation angle, as shown in Fig. 2(c).
Equation (2) can also be applied to the ellipticity $\phi$, and the genuine ellipticity $\xi$ is also plotted in Fig. 2(c).
From this analysis, it can be confirmed that the studied rotationally-stacked woodpile structure can show large optical rotation angle with a small genuine ellipticity in the region 1.2 $\mu$m $<$ $\lambda$ $<$ 2 $\mu$m shaded in Fig. 2(c).
Therefore, clear optical rotation is expected to be observed experimentally in this wavelength range.

\begin{figure}[t]
\centering\includegraphics[width=11cm]{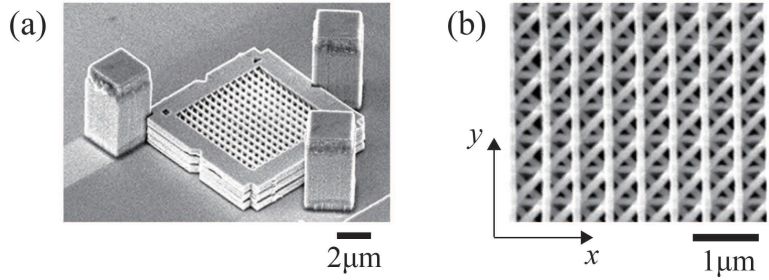}
\caption{
(a) Scanning electron micrograph (SEM) image of the fabricated structure.
Nine plates are stacked using three posts as a guide.
(b) SEM image zoomed on the periodic rods.
The crossing points are aligned by adopting a different rod spacing in the first (third) and second (fourth) layers as described in the main text.
}
\end{figure}

\section{Sample fabrication and experimental setup}
GaAs-based plates were patterned to the same dimensions as those in the numerical simulation by electron beam lithography and dry and wet etching.
Then, the processed plates were stacked by a micro-manipulation technique\cite{Aoki,Aoki2,Aniwat}.
The fabricated structure is shown in Fig. 3.
We characterized the optical rotation by measuring the polarization of laser light transmitted along the helical axis at room temperature.
An infra-red laser with a tunable wavelength range of 1.3 $\mu$m $< \lambda <$ 1.62 $\mu$m is used as a light source.
After synthesizing the polarization of the incident laser with a linear polarizer and wave plates, the laser is focused on the sample normally by a $20\times$ objective lens which has a relatively small numerical aperture (0.45) in order that the depth of focus $\sim$ 5 $\mu$m is larger than the total thickness of the structure.
This enables us to irradiate a plane wave on the sample.
The polarization of the transmitted light is detected using a quarter wave plate, a linear polarizer, and a power meter.

\begin{figure}[t]
\centering\includegraphics[width=12cm]{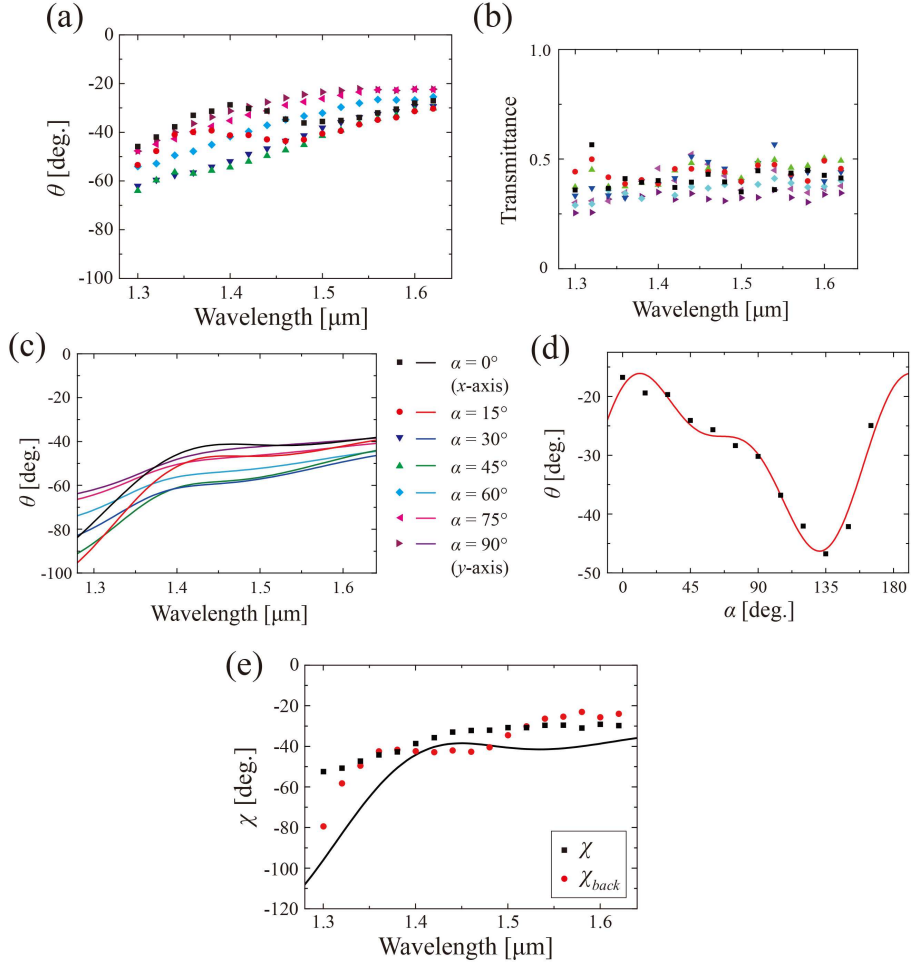}
\caption{
(a) Experimentally obtained dispersion of $\theta$ for various azimuthal angles $\alpha$ of the incident LP.
(b) Transmittance for all measured points in (a).
(c) Numerically obtained dispersion of $\theta$ for various $\alpha$.
The plots are same as those in Fig. 2(a) around $\lambda$ = 1.5 $\mu$m.
The experimental data shows qualitative agreement with the calculated result.
(d) Plot of $\theta$ as a function of $\alpha$ at $\lambda$ = 1.62 $\mu$m. 
A double sinusoidal fitting is performed as mentioned in the main text. 
(e) Dispersion of $\chi$ and $\chi_{back}$ with simulated curve.
The experimental data of $\chi$ for the forward illumination (black dots) which obtained by the fitting for (a) shows qualitative agreement with the calculation (black line) as well as $\chi_{back}$ (red dots) obtained by the backward illumination.
}
\end{figure}

\section{Experimental results and discussions}
Figure 4(a) shows the experimentally obtained dispersion of $\theta$ in the transmitted light for various azimuth angles $\alpha$ of LP in the incident light.
The transmittance through the structure and the substrate for all plots exceeds 0.25 as shown in Fig. 4(b).
In Fig. 4(a), almost all plots show gradual decrease of $|\theta|$ as the wavelength increases.
This tendency is consistent with the calculated results in Fig. 4(c), which is the same plot as Fig. 2(a), but focussing on the 1.28 $\mu$m $<$ $\lambda$ $<$ 1.64 $\mu$m region.
The quantitative difference between the experimental and simulated results is probably because the incident laser is not a complete plane wave with its wave vector slightly tilting for the structure\cite{Konishi2}.
The finite numerical aperture of the objective lens is one of the main reasons for the angle of incidence, slightly modifying the photonic band structure of the 3D PhC. 
Imperfection in the fabrication could be another possible reason.

\begin{figure}[t]
\centering\includegraphics[width=13cm]{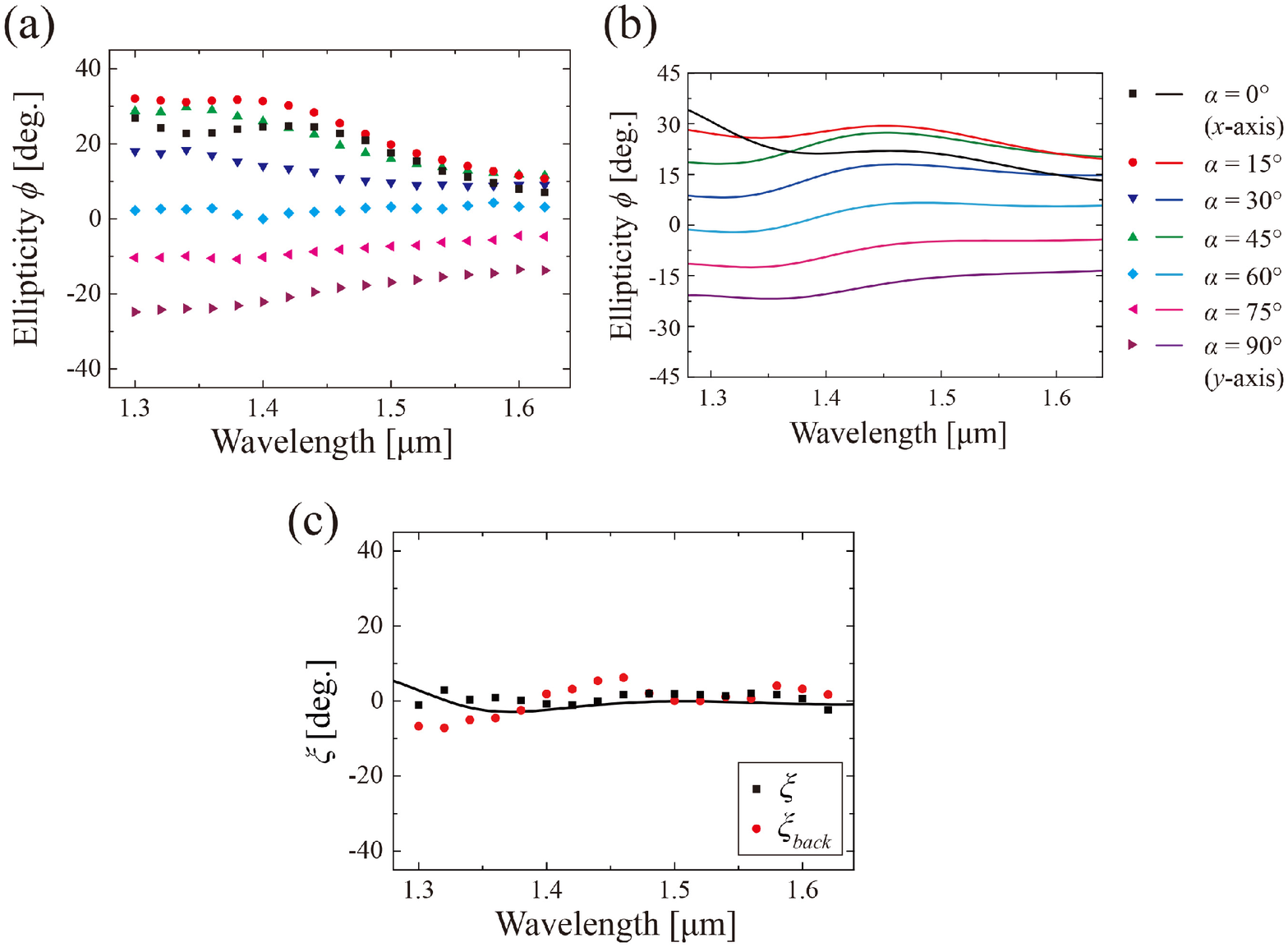}
\caption{
(a) Plot of $\phi$ as a function of the wavelength obtained by the experiment. 
(b) Numerically obtained dispersion of $\phi$ for various azimuthal angle $\alpha$.
The plots are the same as those in Fig. 2(b), around $\lambda$ = 1.5 $\mu$m.
The experimental data shows good agreement with the calculation results. 
(c) Dispersion of $\xi$ and $\xi_{back}$ with a simulated curve.
The experimental results are analyzed and fitted with the equation in the main text.
All three plots are consistent each other and show $\xi$ $\sim$ $0^{\circ}$ throughout the wavelength range.
}
\end{figure}

The experimentally obtained $\theta$ is different for various $\alpha$ at the same wavelength as expected from the numerical calculation.
Using the procedure discussed in Sec. 2, we distinguish between the linear birefringence/dichroism and the optical rotation effect.
As an example, we plot $\theta$ as a function of $\alpha$ at $\lambda$ = 1.62 $\mu$m in Fig. 4(d).
The experimental results can be fitted well with Eq. (2) with the parameters $(\chi, {\rm A_1}, {\rm A_2}, \alpha_1, \alpha_2) = (-29.8^{\circ}, 12.0, 5.67, -28.5^{\circ}, -85.3^{\circ})$.
Note that the term depending on $\alpha$ in Eq. (1) can also be caused by fabrication imperfections and the slightly tilting wave vector of the incident laser in the presense of the linear birefringence/dichroism.
The dispersion of the genuine optical rotation angle $\chi$ is plotted in Fig. 4(e) together with the numerically obtained $\chi$ same as that in Fig. 2(c).
The qualitative tendency shows reasonable agreement with the calculated result.
In the present experiment, we obtained the large optical rotation angle $|\chi| = 52.5^{\circ}$ at $\lambda$ = 1.3 $\mu$m, corresponding to $23.3^{\circ}$ for a single helical unit.
This confirms that the experimentally obtained $\theta$ is dominated by the optical rotation as discussed in Sec. 2.
Comparing $\theta$ with $\chi$, the absolute difference of the effective refractive indices between two orthogonal LPs $|{\Delta}n_L|$ can be estimated as 1/5 of that between two CPs $|{\Delta}n_C|$ at $\lambda$ = 1.3 $\mu$m.

In order to confirm the optical rotation effect we illuminate the laser light from the backside of the sample.
After the fitting by Eq. (2), the optical rotation angle $\chi_{back}$ for the backward illumination is plotted in Fig. 4(e) as red dots.
As expected from the inherent reciprocal nature of optical activity, the plot is almost same as that obtained by the forward illumination, validating the manifestation of the optical rotation.

Figure 5(a) shows the dispersion of the ellipticity $\phi$ obtained by the experiment for the incident light with LP.
The experimental data shows good agreement with the simulated result in Fig. 5(b) same as that in Fig. 2(b) but focussing on the 1.28 $\mu$m $<$ $\lambda$ $<$ 1.64 $\mu$m region.
After fitting with Eq. (2), the genuine ellipticity $\xi$ is plotted in Fig. 5(c) for both simulated and measured results.
We also plot $\xi_{back}$ obtained by the backside illumination.
These three plots are consistent each other, indicating the genuine ellipticity is almost zero in the chiral structure for the wavelength range.

\section{Conclusion}
In summary, we have fabricated a novel rotationally-stacked woodpile structure made from GaAs as a 3D chiral PhC using micro-manipulation, and demonstrate optical activity in the structure.
The dependence of the polarization rotation angle on the input azimuthal angle of LP and its reciprocity confirm the occurrence of optical rotation in the structure.
The experiment demonstrates an optical rotation angle as large as $\sim23^{\circ}$ for a single helical unit at 1.3 $\mu$m wavelength.
To the best of our knowledge, this is the largest optical rotation angle per helical pitch ever reported.
The optical activity in a semiconductor-based 3D chiral PhC could be applicable to CP light emitters, quantum information processing, nano-scale polarization sensors, 3D optical circuits, and even to biological/chemical field for the sorting of enantiomers.

\section*{Acknowledgements}
This work was supported by the Project for Developing Innovation Systems of MEXT and JSPS through its FIRST Program.


\begin{thebibliography}{99}
\bibitem{theory}
In the presense of linear birefringence/dichroism, $\theta$ can exactly be written by a physical approach as the following equation for linearly poralized incident light,
\begin{eqnarray}
\theta = \chi + \frac{1}{2}\arctan\frac{-2p\sin2\alpha-(1-q)\sin4\alpha}{1+q+2p\cos2\alpha+(1-q)\cos4\alpha}.\nonumber
\end{eqnarray}
Here,
\begin{eqnarray}
p = \frac{t_x^2-t_y^2}{t_x^2+t_y^2},\ q = \frac{4t_xt_y\cos\delta}{t_x^2+t_y^2}.\nonumber
\end{eqnarray}
$t_x^2$ and $t_y^2$ are transmitance in the $x$ and $y$ direction, respectively, and $\delta$ is phase difference between $E_x$ and $E_y$ due to linear birefringence.
The term depending on $\alpha$ in this equation shows $\pi$ and $\pi/2$ periodicity, corresponding to the Fourier series of $\sin2\alpha$ in Eq. (1).
In the main text, we derive Eq. (1) by a mathematical approach based on $\pi$ priodicity which gives a simpler explanation than the physical appraoch using the equation above.
\end{thebibliography}
\end{document}